\documentclass[aps,prb,twocolumn,superscriptaddress]{revtex4-1}
\usepackage{graphicx}

\begin{document}

\newcommand{\rmo}{RMn$_{2}$O$_{5}$}
\newcommand{\dymo}{DyMn$_{2}$O$_{5}$}
\newcommand{\mnf}{Mn$^{4+}$}
\newcommand{\mnt}{Mn$^{3+}$}
\newcommand{\dyt}{Dy$^{3+}$}

\bibliographystyle{apsrev4-1}


\title{The Magnetic Structure of DyMn$_{2}$O$_{5}$ Determined by Resonant X-ray Scattering}


\author{G. E. Johnstone}
\email[]{g.johnstone1@physics.ox.ac.uk}

\affiliation{Department of Physics,
        University of Oxford,
        Clarendon Laboratory,
        Parks Road, Oxford, OX1 3PU, United Kingdom}

\author{R. A. Ewings}
\affiliation{ISIS Facility,
        STFC Rutherford Appleton Laboratory,
        Chilton, Didcot, Oxon OX11 0QX, United Kingdom}

\author{R. D. Johnson}
\affiliation{Department of Physics,
        University of Oxford,
        Clarendon Laboratory,
        Parks Road, Oxford, OX1 3PU, United Kingdom}
\affiliation{ISIS Facility,
        STFC Rutherford Appleton Laboratory,
        Chilton, Didcot, Oxon OX11 0QX, United Kingdom}

\author{C. Mazzoli}
\altaffiliation[Current address:]{Dip.to di Fisica, Politecnico di Milano, p.zza L. Da Vinci, 32 ,I-20133 Milano}
\affiliation{ESRF, 6 Rue Jules Horowitz, BP 220, 38043 Grenoble, Cedex 9, France}

\author{H. C. Walker}
\altaffiliation[Current address:]{Deutsches Elektronen-Synchrotron (Hasylab at DESY), 22607 Hamburg, Germany}
\affiliation{ESRF, 6 Rue Jules Horowitz, BP 220, 38043 Grenoble, Cedex 9, France}

\author{A. T. Boothroyd}
\affiliation{Department of Physics,
        University of Oxford,
        Clarendon Laboratory,
        Parks Road, Oxford, OX1 3PU, United Kingdom}


\date{\today}

\begin{abstract}
Resonant magnetic x-ray scattering has been used to investigate the magnetic structure of the magnetoelectric multiferroic \dymo. We have studied the magnetic structure in the ferroelectric phase of this material, which displays the strongest ferroelectric polarisation and magnetodielectric effect of the \rmo\ (where R is a rare earth ion, Y or Bi) family. The magnetic structure observed is similar to that of the other members of the series, but differs in the direction of the ordered moments. In \dymo\ both the Dy and Mn moments lie close to the $b$ axis, whereas in other \rmo\ they lie close to the $a$ axis.
\end{abstract}


\maketitle



\section{Introduction}{\label{intro}}

Magnetoelectric multiferroics have been widely studied for the last half century because of the interesting physics they display \cite{Fiebig2005,Cheong2007,Khomskii2009,Wang2009}, and the prospects for applications based on the mutual control of magnetic order and ferroelectricity \cite{Eerenstein2006}. Much of the recent focus has been on the properties of TbMnO$_{3}$ \cite{Kiruma2003} and similar materials with Tb replaced by another rare earth ion \cite{Goto2004}, and on the \rmo\ series (where R is a rare earth, Y or Bi). The mechanism behind the multiferroicity in these systems is not completely understood, but theories based on symmetry considerations have provided important insights \cite{Mostovoy2006,Bertouras2007}. To construct microscopic models that apply to these systems more information is needed on the details of the various magnetic and electronic phases.


The magnetic properties of many members of the \rmo\ series are well documented through various studies going as far back as the 1960s \cite{Buisson1973,Wilkinson1981}.  There has been a recent resurgence of interest in \rmo\ compounds since the discovery of the magnetically-induced ferroelectricity in these materials \cite{Inomata1996,Hur2004,Hur2004a,Higashiyama2004,Higashiyama2005}. The specific magnetic structures are believed to be important to the magnetoelectric coupling and have been established for many members of the series, usually by neutron diffraction \cite{Wilkinson1981,Blake2005,Vecchini2008,Radaelli2008} or by resonant x-ray scattering \cite{Johnson2008}.

Among the \rmo\ series, \dymo\ displays the largest electric polarization \cite{Fukunaga2010} and magnetodielectric effect\cite{Hur2004a}. Previous studies of the magnetic structure \cite{Wilkinson1981,Blake2005,Ratcliff2005,Ewings2008} have examined the low temperature phase and followed the magnetic ordering wavevector as a function of temperature, but crucially, the magnetic structure in the ferroelectric phase has not been fully determined. Here we address this outstanding issue by reporting the magnetic structure of \dymo\ in the ferroelectric phase as determined by polarized resonant magnetic x-ray scattering (RMXS) and \emph{ab initio} calculations of the resonance spectrum.


The bulk properties of many of the \rmo\ compounds are well documented. They show a ferroelectric polarization along the $b$-axis, approximately in the temperature range $20 \leq T \leq 35$~K.  There exists a transition $T_{N} \approx 40$~K into an incommensurate magnetic phase (HT-ICM), just above the ferroelectric phase. Below the HT-ICM phase, the magnetic structure becomes commensurate and is coexistent with the onset of ferroelectricity (CM-FE phase).  A low temperature incommensurate phase (LT-ICM) appears below 20~K. These phases only have spontaneous magnetic order on the Mn ions, but at lower temperatures still ($T<10$\,K) another phase exists having spontaneous ordering of the magnetic rare earth ions.

In \dymo\, the HT-ICM phase occurs between 42 and 44\,K, and the CM-FE phase exists down to $\sim$15\,K. The LT-ICM phase is observed below 20\,K. \dymo\ displays some differences from the other members of the \rmo\ series. In addition to small variations in the phase transition temperatures, the ordering wavevectors of the incommensurate phases are different, and \dymo\ is unique in having a commensurate structure for the spontaneous order of the rare-earth moments.\cite{Ewings2008}

\begin{figure}[t]
\includegraphics[scale=0.5,angle=0,width=0.5\textwidth]{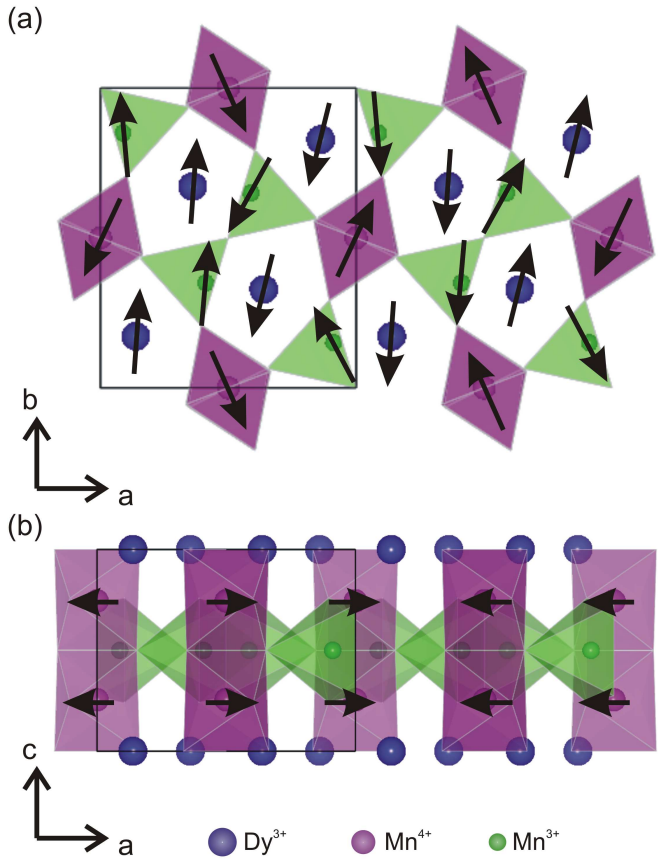}
\centering \caption{\label{fig:struc} The crystal structure and magnetic structure (model A) of \dymo\ viewed along two different crystallographic axes, (a) shows the structure when looking at it along the c-axis and (b) looking at the structure along the b-axis. In both images, the Dy ions are the blue circles, the \mnf\ ions are magenta and are positioned inside the oxygen octahedra and the \mnt\ are shown in green and are positioned inside the oxygen square-based pyramids. The unit cell is shown by the black box. In (b) only the moments on the \mnf sites are shown, demonstrating the interaction between two \mnf moments along the c-axis within a unit cell is ferromagnetic.}
\end{figure}

\dymo\ has a room temperature orthorhombic structure, space group \textit{Pbam}, with lattice parameters $a=7.294$ {\AA}, $b=8.551$ {\AA} and $c=5.688$ {\AA}\cite{Wilkinson1981}. The Mn and O form a structure of Mn$^{4+}$O$_{6}$ octahedra and Mn$^{3+}$O$_{5}$ square-based pyramids, with the Dy ions located in the gaps of the MnO structure, as shown in Fig.\ \ref{fig:struc}. The magnetic structure of \dymo\ is fully established below 8~K \cite{Wilkinson1981,Blake2005}. Neutron diffraction \cite{Ratcliff2005} and RMXS \cite{Ewings2008} experiments were made in the CM-FE phase and confirmed that the magnetic scattering occurs at $(h+1/2,k,l+1/4)$ reflections, but these studies did not determine the magnetic structure in this phase.

The large neutron absorption cross-section of natural Dy limits what information can be obtained by neutron diffraction on the magnetic structure of \dymo.  As an alternative to neutron diffraction, RMXS has two important strengths. Firstly, the resonant enhancement in the scattering when the photon energy is tuned to an atomic transition enhances the otherwise very weak x-ray scattering signal from the magnetic order, and second, a degree of sensitivity to the separate order on the Dy and Mn sublattices can be obtained from the variation in scattering at different absorption edges. These strengths were exploited in the recent RMXS study of magnetism in DyMnO$_3$.\cite{Nandi2008}


\begin{figure}[h]
\includegraphics[scale=0.5,angle=0,width=0.5\textwidth]{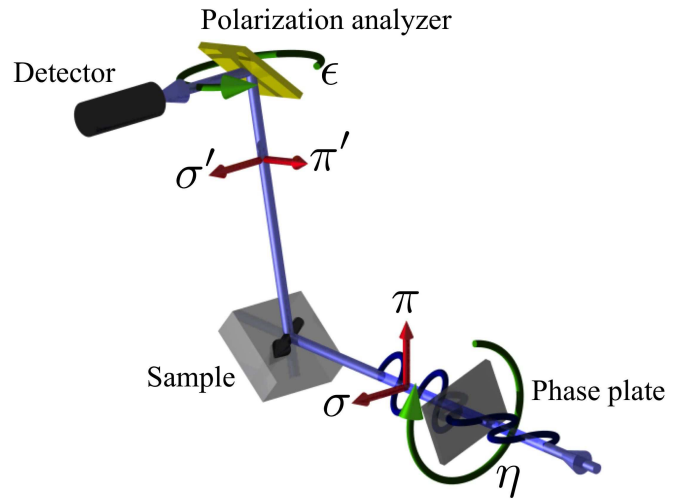}
\centering \caption{\label{fig:scat} The vertical scattering geometry used in this experiment.}
\end{figure}


\section{Experimental Procedure}{\label{exp}}

RMXS was performed at the ESRF, Grenoble. The beamline ID20 \cite{Paolasini2007}, a high-intensity beamline with full polarization analysis, was used with a four-circle diffractometer in vertical scattering geometry, shown in Fig.\ \ref{fig:scat}. A high quality single crystal of \dymo, grown in Oxford {\it et al.} \cite{Wanklyn1972}, was used for this experiment. The crystal was cut so that the $(0,0,1)$ direction was perpendicular to the largest face of the crystal. The crystal was mounted on the cold finger of a closed-cycle refrigerator to allow cooling as low as 12~K.

\begin{figure}[t]
\includegraphics[scale=0.5,angle=0,width=0.5\textwidth]{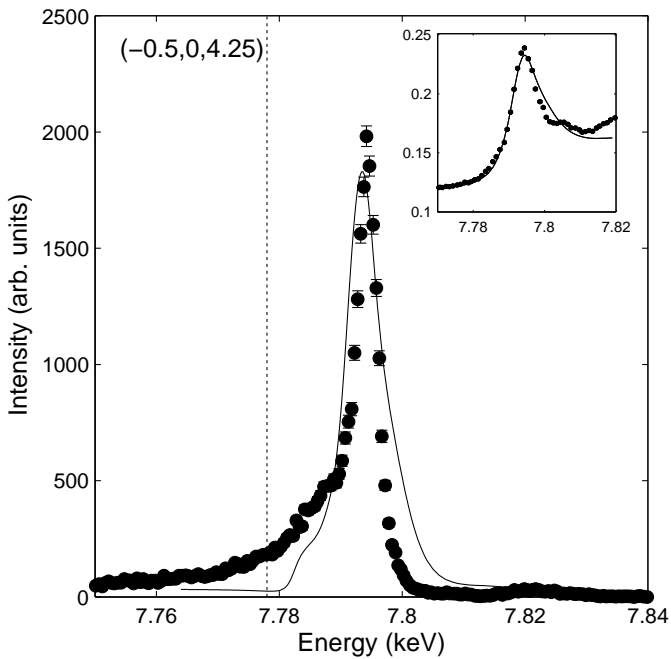}
\centering \caption{\label{fig:fdmnes} A scan of the energy dependence of the (-0.5,0,4.25) magnetic peak, through the resonant Dy L$_{\rm III}$-edge, at 15K in the $\sigma \pi '$ polarisation channel. The measured data is shown with black circles and the line is the energy profile for model A calculated by FDMNES. The inset shows the fluorescence measurement around the same edge, again with the measured data shown with black circles and the FDMNES calculation shown with a black line.}
\end{figure}

Measurements were performed at two x-ray energies, corresponding to two resonance energies in the ions being studied. These were the Dy $L_{\rm III}$-edge ($2p\rightarrow 5d$), and the Mn $K$-edge ($1s\rightarrow 4p$). The measurements were performed at the peak intensity of the resonance spectrum. Fig.\ \ref{fig:fdmnes} shows the energy line shape in the vicinity of the Dy $L_{\rm III}$-edge, measured in the $\sigma\pi'$ polarization channel (which shows a purely magnetic signal) at a temperature of 15\,K, with the x-ray scattering vector fixed at the $(-0.5,0,4.25)$ magnetic Bragg peak. The inset to Fig.\ \ref{fig:fdmnes} shows the fluorescence signal in the same energy range.

Polarization analysis was used to determine the magnetic structure, following the theoretical descriptions of Blume {\it et al.} \cite{Blume1988} and Hill {\it et al.} \cite{Hill1996}. The intensity and polarization of the resonant x-ray diffraction signal is dependent on the magnetic structure. Two different methods of analysing the polarization dependence of the x-ray scattering from the magnetic structure were employed. These are discussed in the following paragraphs.

Azimuthal scans of certain magnetic Bragg reflections were measured. In this type of scan the incoming linear polarization is fixed perpendicular to the scattering plane ($\sigma$ polarisation) and the crystal is rotated around the scattering vector. The analyzer is set so that only scattered x-rays with polarization within the scattering plane ($\pi'$ polarization) are detected. The integrated intensity (measured using a $\theta$-scan) of the magnetic Bragg peak in this $\sigma\pi'$ channel is then obtained as a function of azimuthal angle. This technique has been used to determine details of the magnetic structure in the related materials TbMnO$_{3}$ \cite{Voigt2007,Mannix2007}, HoMn$_{2}$O$_{5}$ \cite{Beutier2008} and TbMn$_{2}$O$_{5}$ \cite{Johnson2008}.

\begin{figure}[t]
\includegraphics[scale=0.5,angle=0,width=0.48\textwidth]{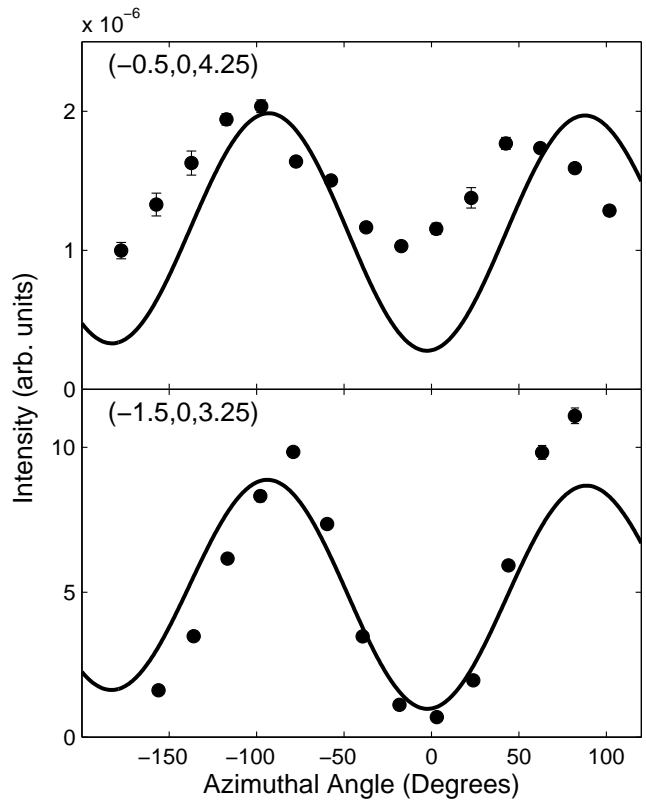}
\centering \caption{\label{fig:dy_azi12} Azimuthal scans performed at the Dy $L_{\rm III}$-edge, at 15K, in the $\sigma \pi'$ polarisation channel. The top panel shows a scan at the  $(-0.5,0,4.25)$ magnetic reflection, and the bottom panel shows the $(-1.5,0,3.25)$ magnetic reflection. The data points are shown with filled circles and the fit to magnetic structure model A is shown by the line.}
\end{figure}

All of the measurements on the CM-FE were conducted at 15~K. In total 5 azimuthal scans were measured, four at the Dy $L_{\rm III}$-edge, at wavevectors $(-0.5,0,4.25)$, $(-1.5,0,3.25)$, $(0.5,1,4.25)$ and $(0.5,2,5.25)$, and one at the Mn $K$-edge, at wavevector $(0.5,1,3.25)$. For all reflections, the azimuth reference vector defining the zero of the azimuth was [0,0,1]. The Dy $L_{\rm III}$-edge scans are shown in Fig.\ \ref{fig:dy_azi12} and \ref{fig:dy_azi34}, and the Mn $K$-edge scan is shown in Fig.\ \ref{fig:mn_azi} together with an energy scan through the weak $K$-edge resonance (insert to Fig.\ \ref{fig:mn_azi}). The data have been corrected for the absorption of the x-ray beam in the sample, which varies with azimuthal angle.\cite{Beutier2008} This correction is essential because the absorption varies strongly with the angle that the incoming and outgoing x-ray beams make with the crystal surface.



The alternative method is to use full linear polarization analysis (FLPA), in which the incoming x-ray polarization is varied by use of phase-plates, while keeping the sample in a fixed position.\cite{Mazzoli2007,Johnson2008} This has two advantages over the azimuthal scan. First the attenuation does not vary during the measurement, and second, the sample does not move so the same part of the sample is measured. The sample was aligned on the $(-0.5,0,4.25)$ reflection at 15\,K, again using the resonance at the Dy $L_{\rm III}$-edge. The azimuthal angle the scan was performed at was $-97.5^{\circ}$, the maximum of the intensity in the azimuthal scan at this wavevector. The angle of the incoming polarization, $\eta$, is set so that $\eta=0$ corresponds to $\sigma$ polarised light (and therefore $\pi$ polarised light is at $\eta=90^{\circ}$). For each value of incoming polarization measured, a FLPA of the outgoing beam was performed to obtain the values of the first two Stokes parameters, $P_{1}$ and $P_{2}$. The parameter $P_{1}$ describes the polarization perpendicular and parallel to the scattering plane and $P_{2}$ the polarization at $\pm 45^{\circ}$ to the scattering plane. The two parameters therefore characterise the linear polarization, with $P_{lin}=\sqrt{P_{1}^{2}+P_{2}^{2}}$. There is a third parameter, $P_{3}$, which is related to the circular polarization. Fig.\ \ref{fig:dy_stokes} shows $P_{1}$ and $P_{2}$ obtained this way as a function of $\eta$.

\section{Results}{\label{res}}


To determine the magnetic structure of \dymo, we treated the analysis in three parts. The magnetic ordering on the Dy$^{3+}$ and Mn$^{4+}$ sites was determined from a fit to the four azimuthal scans and the FLPA $\eta$ scan performed at the Dy $L_{\rm III}$ edge. Having obtained the Mn$^{4+}$ magnetic structure, we determined the ordering on the Mn$^{3+}$ ions from a fit to the azimuthal scan at the Mn $K$ edge. Finally, we  confirmed the validity of the results by comparing the energy scans over the Dy $L_{\rm III}$ edge with {\it ab initio} calculations. The procedure is explained in the following paragraphs.

X-ray magnetic scattering at the Dy $L_{\rm III}$ edge probes the empty Dy $5d$ states. As has been discussed previously,\cite{Johnson2008} the magnetic polarization of the $5d$ states arises not only from the localized Dy $4f$ moments but also from the \mnf\ moments due to the overlap between the extended Dy $5d$ states and the \mnf\ $3d$ orbitals. The \mnt\ ions are thought to have a negligible contribution at the Dy $L_{\rm III}$ edge, as they are situated further from the Dy ions in the crystal structure.

Previous work has shown that at low temperatures the spontaneous magnetic order of the rare earth sublattice does not always have the same symmetry as that of the Mn sublattices \cite{Johnson2008,Johnson2011}. Following work on TbMn$_2$O$_5$ and HoMn$_2$O$_5$ \cite{Vecchini2008,Johnson2008}, we constrained the direction of the \dyt\ moments on pairs of sites to be the same. Specifically, sites (1,2) have one direction and sites (3,4) have another direction (see Table \ref{tab:mag} for the positions of the sites), and we constrained the size of all the \dyt\ moments to be the same. For the Mn magnetic order we assumed the same structure as found many times previously \cite{Radaelli2008}. This assumption has the \mnf\ moments linked ferromagnetically along the c-axis within a unit cell. The \mnf\ ions at $x=0$ and $x=0.5$ are linked by a $b$-glide. The \mnt\ structure has two Mn ions linked by a $b$-glide and the other two linked by a $b'$-glide. Initially we neglected the component of the ordered moments along the $c$ direction, which is reasonable given that the magnetic structures of other \rmo\ compounds have only a small $c$-axis component.

\begin{figure}[t]
\includegraphics[scale=0.5,angle=0,width=0.48\textwidth]{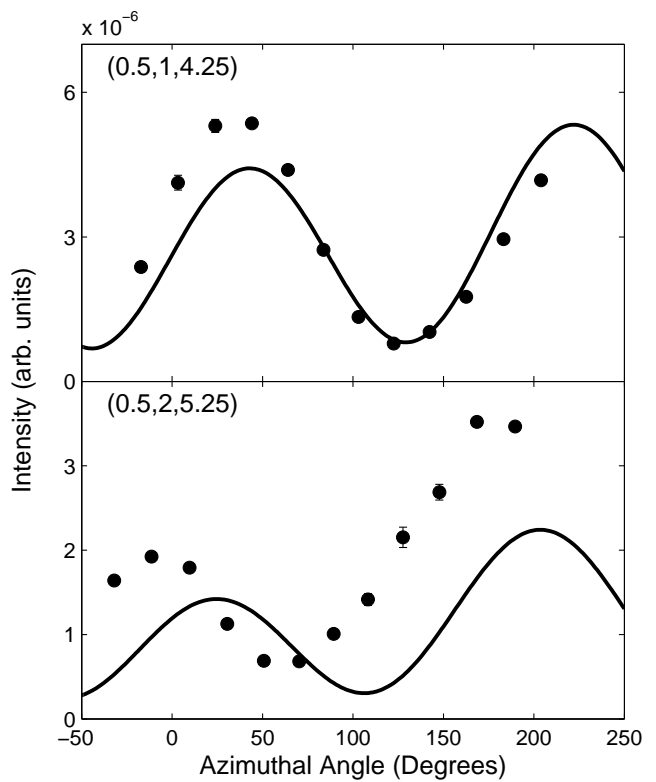}
\centering \caption{\label{fig:dy_azi34} Azimuthal scans performed at the Dy $L_{\rm III}$-edge, at 15K, in the $\sigma \pi'$ polarization channel. The top panel shows a scan at the  $(0.5,1,4.25)$ magnetic reflection, and the bottom panel shows the $(0.5,2,5.25)$ magnetic reflection. The data points are shown with filled circles and the fit to magnetic structure model A is shown by the line.}
\end{figure}

\begin{figure}[t]
\includegraphics[scale=0.5,angle=0,width=0.5\textwidth]{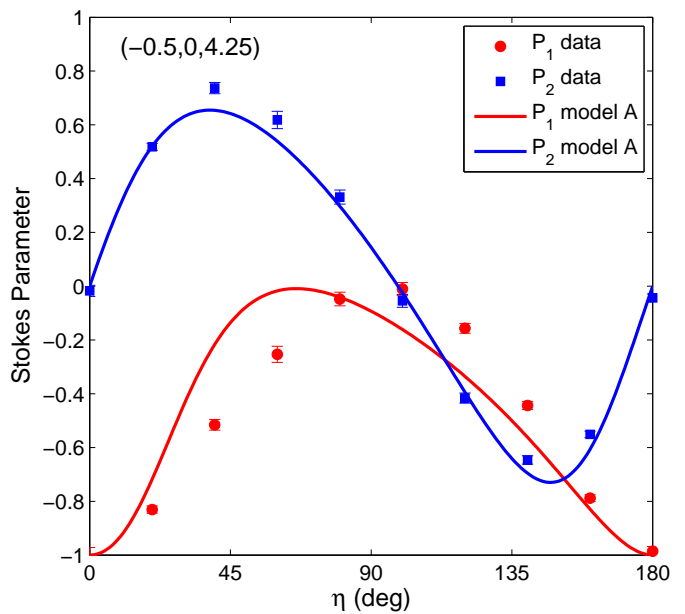}
\centering \caption{\label{fig:dy_stokes} A full linear polarization analysis measurement of the Stokes parameters $P_1$ and $P_2$ as a function of incoming polarization ($\eta$). The data were collected at the Dy ${L}_{\rm III}$-edge at a temperature of 15\,K. The X-ray scattering wavevector is fixed at $(-0.5,0,4.25)$ with the angle of the azimuth at -97.5$^{\circ}$, the maximum of the azimuthal scan at this wavevector. The symbols represent the experimental data, and the lines are calculated from the magnetic structure model A.}
\end{figure}

The fit to the data at the Dy $L_{\rm III}$-edge was performed simultaneously to the four azimuthal scans and the FLPA $\eta$ scan (Figs.~\ref{fig:dy_azi12}, \ref{fig:dy_azi34} and \ref{fig:dy_stokes}). The free parameters were the three in-plane angles of the ordered moments (one for each of the inequivalent Dy sites, and one for the \mnf site) and the fraction $\alpha$ of the \mnf contribution to the signal (the remainder coming from the \dyt).  The best fit was achieved with $\alpha= 0.64(7)$.  We note that it is not possible to obtain a good fit to the Dy $L_{\rm III}$-edge data allowing only \dyt\ ions in the model ($\alpha=0$), nor is it possible using only \mnf ions in the model ($\alpha=1$). This demonstrates that the signal at the  Dy $L_{\rm III}$-edge arises from both Dy and Mn magnetism.


\begin{figure}[t]
\includegraphics[scale=0.5,angle=0,width=0.5\textwidth]{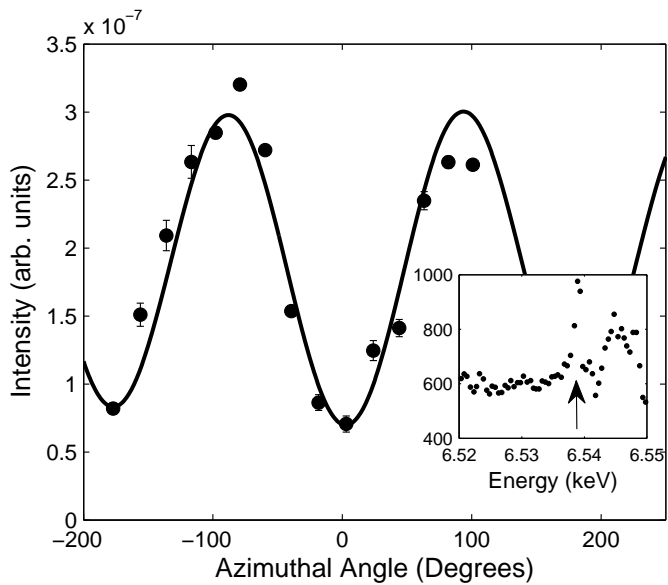}
\centering \caption{\label{fig:mn_azi} Azimuthal scan performed at the Mn $K$-edge, at 15K, in the $\sigma \pi'$ polarization channel. The wavevector is $(0.5,1,3.25)$. The experimental data are shown with filled circles, and the fit to magnetic structure model A is shown by the line. The insert shows the energy line shape of the Mn $K$-edge resonance at the same wavevector and temperature, measured at an azimuthal angle of -79$^{\circ}$. The resonance peak is marked by the black arrow.}
\end{figure}


At the Mn $K$-edge, the empty $4p$ states of the Mn ions are probed, but the magnetism is carried by the 3$d$ electrons. The Mn $K$-edge scan is sensitive to the \mnf\ and \mnt\ moments via an intra-atomic Coulomb interaction. We assume the contribution from the Dy moments is negligible because the Dy 4$f$ and Mn $4p$ states are highly localized. We fitted the Mn $K$-edge azimuthal scan to a model for the magnetic ordering of the \mnt\ ions assuming the magnetic structure of the \mnf\ ions determined previously from the data at the Dy $L_{\rm III}$-edge. In principle, it should be possible to distinguish between \mnt\ and \mnf\ signals using the characteristic energy line shape of the Mn K-edge resonance. Unfortunately, the K-edge enhancement is very weak, and the data are not good enough for this type of analysis (Fig.\ \ref{fig:mn_azi} insert). Instead, we determined the ratio $\beta$ of the contribution from \mnt\ to \mnf\ by the same method as described in the previous paragraph for $\alpha$. The best fit is shown in Fig.\ \ref{fig:mn_azi} and had $\beta = 0.55(2)$.

Using this method, several equally good fits of the azimuthal and FLPA $\eta$ scans were found. However, all but one of the models could be dismissed from analysis of Dy $L_{\rm III}$-edge resonant energy spectrum. To differentiate between these magnetic structure models, \textit{ab-initio} calculations of the Dy ${L}_{\rm III}$-edge resonance at the $(-0.5,0,4.25)$ reflection were performed. The FDMNES code was employed \cite{Joly2001}, which performs a monoelectronic, cluster-based multiple scattering calculation, well suited to the prediction of resonant spectra that involve excitations into delocalized (rare-earth $5d$) states. A 256 atom, $2a \times b \times 4c$ supercell (the magnetic unit cell) was defined, containing 32 Dy atoms, located on 8 symmetry inequivalent sites.  A cluster radius of 4.5~$\mathrm{\AA}$ was selected, above which no improvement in the calculation accuracy was observed. First, the absolute energy scale was calibrated by calculating the Dy ${L}_{\rm III}$ absorption spectra and comparing to the experimentally measured fluorescence, shown in the inset of Fig.\ \ref{fig:fdmnes}. The Fermi energy was then set at $E_{\rm F}= 7.778$~keV, giving a low energy cut-off. The resonance was calculated with both E1--E1 and E2--E2 excitation channels allowed. The E2--E2 excitation, however, gave a negligible contribution to the resonant intensity and was therefore omitted from the final calculations. By comparing calculations based on the possible magnetic structure models it was clear that only one model comes close to reproducing the resonant energy spectrum, which is shown in Fig.\ \ref{fig:fdmnes}.

The magnetic structure which is consistent with the azimuth scans, FLPA scans, and resonance lineshape is presented in table \ref{tab:mag} and depicted in Fig.\ \ref{fig:struc}. This model of the in-plane magnetic structure will hereafter be referred to as model A. As mentioned earlier, we only fitted the in-plane components of the magnetic moments. The directions of the moments are therefore specified by their angle $\phi$ from the $a$-axis.  The layers stack with alternating ferromagnetic and antiferromagnetic coupling along the $c$ axis. This gives the $\uparrow \uparrow \downarrow \downarrow$ propagation of the magnetic structure along the c-axis that is observed in other \rmo\ materials. 

The fact that the temperature at which these experiments were performed (15\,K) is above the spontaneous ordering temperature for Dy (8\,K) suggests that any ordered magnetic moments on the Dy sites are induced by the magnetic order of the Mn sublattice. In this scenario, it is expected that the ordering on the Dy sites should follow the symmetry of the Mn magnetic structure. However, this symmetry constraint is not applied in model A. Therefore, we tested a second model (model B) with the moments on the Dy ions following the same symmetry as the Mn moments.

In model B, the Dy moments are still considered in pairs, but rather than being parallel, as in model A, the moments in each pair are related by either a $b$-glide or $b'$-glide. Under these constraints the best fit to the data at the Dy L$_{\rm III}$ edge is found when $\alpha = 0$, meaning that only the magnetic moments at the Dy sites contribute to the signal. The calculated azimuthal and FLPA $\eta$ scans for this fit are very similar to the ones presented in figures \ref{fig:dy_azi12}, \ref{fig:dy_azi34} and \ref{fig:dy_stokes}, and are therefore not shown here. The value of the reduced $\chi ^ {2}$ is about $3\%$ higher for the best fit to model B when compared to the best fit to model A. An FDMNES simulation (not shown here) assuming the best-fit model B structure is consistent with the measured Dy ${L}_{\rm III}$ edge resonance at the $(-0.5,0,4.25)$ reflection. Fig. \ref{fig:dy_altstruc} depicts the model B magnetic structure.

The experimental geometry used meant that we had limited sensitivity to the component of the magnetic moment parallel to the $c$-axis. Test models in which a $c$-axis component was added by hand confirmed this insensitivity, which is caused by the relatively small angles that the scattering vectors at each of the reflections studied make with the $c$-axis. We cannot, however, rule out a small $c$-axis component to the ordered moments, but as noted earlier, in other \rmo\ compounds the moments lie predominantly in the $ab$-plane.

\begin{table}[t]
\caption{\label{tab:mag} The magnetic structure (model A) of \dymo\ at 15\,K, in the commensurate antiferromagnetic and ferroelectric phase. The positions of the ions have been determined elsewhere \cite{Wilkinson1981}. The measured moment directions lie in the $ab$-plane and are specified by the angle $\phi$ from the $a$-axis. The numbers in brackets in the final column are the uncertainties in the last digits obtained from the fits.}
\begin{ruledtabular}
\begin{tabular}{c c c c c}
Atom     & $x$      & $y$      & $z$      & $\phi$ (deg)  \\
\hline
\mnf\ (1) & 0      & 0.5    & 0.2521 & 245(3)         \\
\mnf\ (2) & 0      & 0.5    & 0.7479 & 245(3)          \\
\mnf\ (3) & 0.5    & 0      & 0.2521 & 294(3)      \\
\mnf\ (4) & 0.5    & 0      & 0.7479 & 294(3)      \\
\mnt\ (1) & 0.0759 & 0.8447 & 0.5    & 95.2(7)     \\
\mnt\ (2) & 0.4241 & 0.3447 & 0.5    & 85.8(7)      \\
\mnt\ (3) & 0.5759 & 0.6553 & 0.5    & 241.2(2)      \\
\mnt\ (4) & 0.9241 & 0.1553 & 0.5    & 118.8(2)      \\
\dyt\ (1) & 0.1389 & 0.1729 & 0      & 86(10)      \\
\dyt\ (2) & 0.3611 & 0.6729 & 0      & 86(10)      \\
\dyt\ (3) & 0.6389 & 0.3271 & 0      & 256(10)     \\
\dyt\ (4) & 0.8611 & 0.8271 & 0      & 256(10)     \\
\end{tabular}
\end{ruledtabular}
\end{table}

\section{Discussion}{\label{dis}}

\begin{figure}[t]
\includegraphics[scale=0.5,angle=0,width=0.5\textwidth]{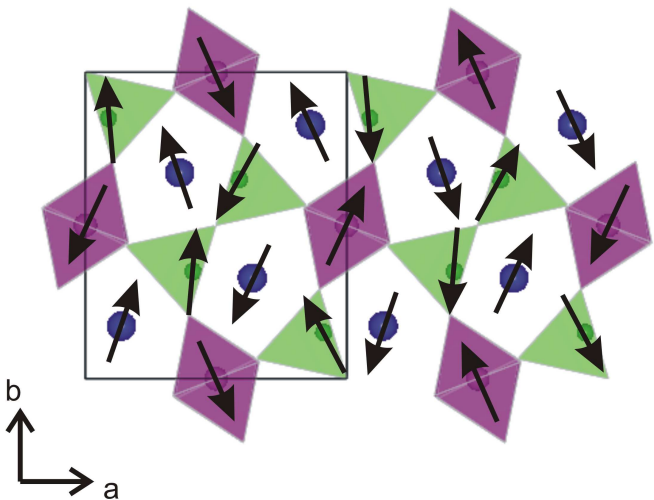}
\centering \caption{\label{fig:dy_altstruc} Magnetic structure model B of \dymo, in which the Dy magnetic structure is constrained to have the same symmetry as the Mn magnetic structure.}
\end{figure}

The magnetic structure of the Mn ions, found here for the ferroelectric phase, is similar to the magnetic structure of the Mn ions found in \dymo\ in the Dy-ordered phase by previous studies \cite{Blake2005,Wilkinson1981}. The structure shows approximately antiferromagnetic chains running parallel to the $a$-axis, a common feature of the \rmo\ series that agrees with the magnetic exchanges discussed elsewhere \cite{Chapon2004,Blake2005}. However, one distinct difference between \dymo\ and the other members of the \rmo\ series is that in the ferroelectric phase measured here, the magnetic moments are aligned close to parallel with the $b$-axis. This is in contrast to the other members of the series which tend to have the magnetic moments approximately parallel to the $a$-axis.

The best fit model we obtained (model A) breaks the space group symmetry on the Dy site. In this model, the Dy moments occur in ferromagnetically-aligned pairs. This is similar to the rare-earth magnetic ordering in TbMn$_2$O$_5$, \cite{Johnson2008}, but in contrast to that in HoMn$_2$O$_5$ \cite{Vecchini2008}. However, at 15~K, the material is above the spontaneous magnetic ordering transition temperature of the Dy ions, so at this temperature it would be expected that any static order of the Dy moments would be induced by coupling to the Mn sublattice and would therefore have the same symmetry as the Mn magnetic order. This is the case for model B (by construction) but not for model A. Therefore, if model A describes the correct structure, then the implication is that the Dy--Dy magnetic interactions remain a sufficiently strong influence at 15~K to prevent the Dy magnetic order from adopting the same symmetry as the Mn magnetic order.

The differences between the Dy magnetic structures of model A and model B are actually relatively small. In model A, the Dy moment on site 4 is parallel to that on site 3, whereas in model B the moments on sites 3 and 4 are approximately antiparallel (see Figs.~ \ref{fig:struc}(a) and \ref{fig:dy_altstruc}). The other Dy moments are found to have only a slight difference in their angle to the $a$-axis between the two models. In both models the Dy moments are close to parallel with the $b$-axis, again in contrast to the observations of other members of the \rmo\ series.


There are a few noteworthy features of the fit itself. The fit successfully describes three out of four of the Dy L$_{\rm III}$-edge azimuthal scans, but does not adequately describe the azimuthal scan at $(0.5,2,5.25)$.  When treating this scan independently of all other scans, we could find no model that describes the observed azimuthal dependence. One possibility is that there is a systematic error due to sample absorption. However, calculations indicate that the variation in absorption is not significantly greater than at the other reflections.




In the FLPA $\eta$-scan, around the point $\eta=90^{\circ}$, $P_{1}^{2}$ and $P_{2}^{2}$ both tend to zero. The magnitude of the Stokes vector $(P_{1},P_{2},P_{3})$ should be equal to one for polarized light. It is obvious that at $\eta=90^{\circ}$, $P_{lin} \neq 1$ (Fig.\ \ref{fig:dy_stokes}). The data suggests that the beam is either depolarized or there is a circular component to the polarization. This effect has been observed before and was explained by a non-zero $P_{3}$ and therefore a circular component to the polarization of the outgoing beam.\cite{Mazzoli2007,Bland2009,Bland2010}

\section{Conclusion}{\label{conclusion}}

In this work we have investigated the magnetic structure in the multiferroic phase of \dymo, which has not been considered in detail before now. We have found two alternative but similar models which provide a good description of the data.  Although the symmetry of the rare-earth and Mn magnetic structures in \dymo\ is the same as that found in other \rmo\ compounds, a key difference is that in \dymo\ the Dy and Mn moments point in a direction close to the $b$ axis, whereas in other \rmo\ compounds the moments are close to the $a$ axis. Given the strong coupling between magnetism and ferroelectricity in the \rmo\ series, it is likely that this difference plays a role in causing the exceptionally large electric polarisation in \dymo.


\section{Acknowledgments}{\label{acknowledgments}}

The single crystal sample used in these experiments was prepared by B. M. Wanklyn. GEJ is grateful to the Engineering and Physical Sciences Research Council and the Science and Technology Facilities Council for financial support. The authors would like to J. Payne for his assistance during the experiment.

\bibliography{refs_dymo_aps}

\end{document}